\documentclass[12pt]{article}
\usepackage{graphicx}
\usepackage{color}
\begin{document}

\title{A relativistic study of Bessel beams}
\author{S. Hacyan and R. J\'auregui}

\maketitle
\begin{center}

{\it  Instituto de F\'{\i}sica,} {\it Universidad Nacional
Aut\'onoma de M\'exico,}

{\it Apdo. Postal 20-364, M\'exico D. F. 01000, Mexico.}

\end{center}
\vskip0.5cm

We present a fully relativistic analysis of Bessel beams revealing
some noteworthy features that are not explicit in the standard
description. It is shown that there is a reference frame in which
the field takes a particularly simple form, the wave appearing to
rotate in circles. The concepts of polarization and angular
momentum for Bessel beams is also reanalyzed.



\newpage
Bessel beams are solutions of Maxwell's equations with important
properties: they propagate with an intensity pattern that remains
invariant along a given axis \cite{Durnin1}, and carry an angular
momentum that is not due to their polarization state \cite{OAM}.
Their experimental realization in laboratories has attracted much
attention in recent years \cite{exp}.

The aim of the present paper is to study the general properties of
Bessel beams within a fully relativistic framework. We show that,
contrary to the case of a plane wave, there is a privileged
reference frame in which the linear propagation is eliminated and
the wave propagates circularly. This particular frame corresponds,
in a certain sense, to an {\it antiparaxial} limit. A general
implication is that Maxwell's equations admit as solutions
electromagnetic waves that propagate in circles.

Our starting point is the definition of a Bessel electromagnetic
wave in terms of Hertz potentials $\Pi_1$ and $\Pi_2$
\cite{nisbet}. In cylindrical coordinates $\{\rho , \phi , z \}$,
the electromagnetic potentials are given by \cite{jh}
\begin{equation}
\Phi = - \frac{\partial}{\partial z}\Pi_1,
\end{equation}
\begin{equation}
{\bf A} = \Big\{ \frac{1}{\rho}\frac{\partial}{\partial \phi}\Pi_2
, -\frac{\partial}{\partial \rho}\Pi_2 , \frac{\partial}{\partial
t}\Pi_1 \Big\},
\end{equation}
(in units with $c=1$), and both Hertz potentials $\Pi_i$ ($i=1,2$)
satisfy the equation:
\begin{equation}
- \frac{\partial^2}{\partial t^2}\Pi_i + \frac{1}{\rho}
\frac{\partial}{\partial \rho} \Big(\rho \frac{\partial}{\partial
\rho} \Pi_i \Big) + \frac{1}{\rho^2} \frac{\partial^2}{\partial
\phi^2}\Pi_i + \frac{\partial^2}{\partial z^2}\Pi_i = 0.
\end{equation}
Any solution of this equation regular at the origin can be written
as a linear combination of the functions
\begin{equation}
\Pi_1 = \frac{{\cal B}_K}{k_\bot^2} J_m(k_{\bot} \rho)
\exp\{-i\omega t + ik_zz + im\phi\},
\end{equation}
and
\begin{equation}
\Pi_2 = \frac{{\cal E}_K}{k_{\bot}^2} J_m(k_{\bot} \rho)
\exp\{-i\omega t + ik_zz + im\phi\},
\end{equation}
where $J_m$ is the Bessel function of order $m$, ${\cal E}_K$ and
${\cal B}_K$ are constants, and $k_{\bot}=\sqrt{\omega^2 - k_z^2}$
is the transverse wave-number; here and in the following, the set
of numbers $\{k_{\bot},m,k_z\}$ will be denoted with the generic
symbol $K$.  An electromagnetic mode is associated to each Hertz
potential, giving rise to transverse magnetic (TM) and electric
(TE) modes respectively. Using a Cartesian basis, it follows that
the most general superposition of these waves is given by
\cite{comment}
\begin{eqnarray}
{\bf E}_K  = \frac{1}{2 k_\bot}
 e^{-i\omega t + i k_z z} &\Big[&(\omega {\cal E}_K + i k_z {\cal B}_K)
J_{m-1} (k_\bot \rho) e^{i (m-1) \phi} (\hat{{\bf e}}_x + i
\hat{{\bf e}}_y) \nonumber\\ &+& (\omega {\cal E}_K -i k_z {\cal
B}_K ) J_{m+1} (k_\bot \rho) e^{i (m+1) \phi} (\hat{{\bf e}}_x - i
\hat{{\bf e}}_y)~~\Big]\nonumber\\
&+& e^{-i\omega t + i k_z z} {\cal B}_K J_{m} (k_\bot \rho) e^{i m
\phi} \hat{{\bf e}}_z,
\end{eqnarray}
and
\begin{eqnarray}
{\bf B}_K  = \frac{1}{2 k_\bot}
 e^{-i\omega t + i k_z z} &\Big[&(i k_z {\cal E}_K - \omega {\cal B}_K)
J_{m-1} (k_\bot \rho) e^{i (m-1) \phi} (\hat{{\bf e}}_x + i
\hat{{\bf e}}_y) \nonumber\\ &-& (i k_z {\cal E}_K + \omega {\cal
B}_K) J_{m+1} (k_\bot \rho) e^{i (m+1) \phi} (\hat{{\bf e}}_x - i
\hat{{\bf e}}_y)~~\Big]\nonumber\\
&+& e^{-i\omega t + i k_z z} {\cal E}_K J_{m} (k_\bot \rho) e^{i m
\phi}  \hat{{\bf e}}_z~.
\end{eqnarray}
Clearly the electric an magnetic fields are interchanged under a
duality transformation ${\cal B}_K \rightarrow {\cal E}_K$ and
${\cal E}_K\rightarrow -{\cal B}_K$. A compact way to write the
above expressions is:
\begin{eqnarray}
{\bf E}_K  &\pm& i {\bf B}_K = ({\cal E}_K \mp i{\cal B}_K)
e^{-i\omega t + i k_z z} \Big[~ \frac{\omega \mp k_z}{2 k_{\bot}}
~J_{m-1} ~ e^{i (m-1) \phi} (\hat{{\bf e}}_x + i
\hat{{\bf e}}_y) \nonumber\\
&+& \frac{\omega \pm k_z}{2 k_{\bot}} ~J_{m+1} ~ e^{i (m+1) \phi}
(\hat{{\bf e}}_x - i \hat{{\bf e}}_y) \pm iJ_{m} ~e^{i m \phi}
\hat{{\bf e}}_z \Big]
\end{eqnarray}
(here and in the following, the argument of all Bessel functions
is $k_\bot \rho$).

The complex Lorentz invariant of the electromagnetic field follows
from the above expressions:
$$
({\bf E}_K + i {\bf B}_K) \cdot ({\bf E}_K^* + i {\bf B}_K^* ) =
$$
\begin{equation}
\frac{1}{2} ({\cal E}_K - i{\cal B}_K)({\cal E}_K^* - i{\cal B}_K^*) \Big(
 | J_{m+1} |^2 +  | J_{m-1} |^2  - 2| J_{m} |~^2
\Big),\label{eq:complexinv}
\end{equation}
and also
$$
(\Re{\bf E}_K)^2 - (\Re{\bf B}_K)^2 + 2 i (\Re{\bf E}_K)\cdot
(\Re{\bf B}_K) =
$$
$$
\frac{1}{4}\Big[({\cal E}_K - i{\cal B}_K) e^{-i\omega t + i k_z z+im\phi} + ({\cal
E}_K^* - i{\cal B}_K^*) e^{i\omega t - i k_z z-im\phi}\Big]^2  ( J_{m+1}
 J_{m-1}  -  J_{m}^2 )
$$
\begin{equation}
+\frac{1}{4}({\cal E}_K - i{\cal B}_K)({\cal E}_K^* - i{\cal
B}_K^*)(J_{m+1}-J_{m-1})^2 , \label{eq:realinv}
\end{equation}
where $\Re{\bf E}$ and $\Re{\bf B}$ are the real parts of the
electric and magnetic fields, that is, the physically measurable
quantities. Notice that, unlike the case of a plane wave, the
Lorentz invariant $\Re{\bf E}_K\cdot \Re{\bf B}_K$ can be
different from zero. From Maxwell equations, this implies that
\begin{equation}
\Re{\bf E}_K\cdot\nabla \times \Re{\bf E}_K \neq 0
\label{eq:north}
\end{equation}
for monochromatic waves. A similar expression follows for the magnetic field.
Eq.~(\ref{eq:north}) can be satisfied only if there is no scalar field $g({\bf r})$
such that $\nabla \times (g\Re{\bf E}_K) =0$. Thus the rotational behavior of the
electric and magnetic field cannot be removed continuously: this is a manifestation
of the fact that Bessel beams have optical vortices.

There are two particularly important superpositions of Bessel
modes: the so-called right and left polarized states, given by the
conditions
\begin{equation}
\omega {\cal E}_K \pm i k_z {\cal B}_K = 0 \label{eq:RL},
\end{equation}
and the states for which
\begin{equation}
{\cal E}_K \pm i {\cal B}_K=0 . \label{eq:pm}
\end{equation}
The first condition corresponds to a wave whose electric field has components along
either $\hat{{\bf e}}_x + i \hat{{\bf e}}_y$ or $\hat{{\bf e}}_x - i \hat{{\bf
e}}_y$, and its $z$ component is negligible in the paraxial approximation; however,
the corresponding magnetic field in this case has a rather complicated structure. The
second definition appears more naturally within a quantum optical framework: it
corresponds to a basis of orthonormal modes in which the operators for the energy,
the helicity, and the z-components of linear and orbital angular momentum are all
simultaneously diagonal \cite{jh}. Eqs.~(\ref{eq:RL}) or Eqs.~(\ref{eq:pm}) are
equivalent in the paraxial approximation.

For circularly polarized beams in the sense of Eq.~(\ref{eq:pm}),
the complex Lorentz invariant vanishes, in close analogy with a
standard plane wave of arbitrary polarization, while the Lorentz
invariant given by Eq.~(\ref{eq:realinv}) takes a simple form but
it is not strictly zero. This means that the {\it temporal
averages} of $\vert\Re{\bf E}_K \vert^2 - \vert\Re{\bf B}_K
\vert^2$ and $\Re{\bf E}_K\cdot\Re{\bf B}_K$ are zero, although
the corresponding values at a given time do not vanish.

At this point, we notice that there is a particular reference frame in which TE and
TM Bessel modes take a simpler form. This is the frame moving along the $z$ axis with
velocity $v=k_z/\omega$ and Lorentz factor $\gamma = \omega/k_{\bot}$. The standard
Lorentz transformation to this new frame is
$$
t^{\prime} = \frac{1}{k_{\bot}}(\omega t - k_zz)
$$
\begin{equation}
z^{\prime} =\frac{1}{k_{\bot}}(\omega z - k_z t),
\end{equation}
and the same transformation holds for the $t$ and $z$ components
of the potential $A^{\mu}$. Thus the transformed electromagnetic
field does not depend on the coordinate $z^\prime$, since
$k_z^\prime =0$. Changing to this frame is equivalent to taking an
$antiparaxial$ limit. If the mode satisfies the paraxial
approximation in the laboratory frame, the above Lorentz
transformation involves an ultra relativistic velocity, $v\sim 1$.

A direct calculation shows that the scalar potential $\Phi = 0$ in the new frame, and
therefore the Coulomb gauge is satisfied directly. As for the vector potential, it
takes the form:
\begin{equation}
{\bf A}^{\prime} = -\frac{i}{k_{\bot}} ({\cal E}_K {\bf u}_K +
{\cal B}_K  {\bf v}_K ),
\end{equation}
where
\begin{eqnarray}
{\bf u}_K &=& \Big[ \frac{m}{k_{\bot}\rho} J_m ~ {\bf e}_{\rho} +
i J^{\prime}_m  ~{\bf e}_{\phi}\Big]
e^{-i k_{\bot}t ^{\prime}+ i m\phi}\nonumber\\
&=& \frac{1}{2} \Big[ J_{m-1} ~e^{i (m-1) \phi} (\hat{{\bf e}}_x +
i \hat{{\bf e}}_y) + J_{m+1} ~e^{i (m+1) \phi} (\hat{{\bf e}}_x -
i \hat{{\bf e}}_y)\Big]e^{-i k_{\bot}t ^{\prime}}
\end{eqnarray}
and
\begin{equation}
{\bf v}_K = J_m ~e^{-i k_{\bot}t ^{\prime}+ i m\phi} {\bf e}_z .
\end{equation}

The electric and magnetic fields are
\begin{equation}
{\bf E}^{\prime}=  {\cal E}_K {\bf u}_K + {\cal B}_K {\bf v}_K
\end{equation}
and
\begin{equation}
{\bf B}^{\prime}= - {\cal B}_K {\bf u}_K + {\cal E}_K {\bf v}_K .
\end{equation}
In the antiparaxial reference frame, the magnetic (electric) field
of a transverse electric (magnetic) mode is parallel to the $z$
axis.

 The standard definition, Eq.~(\ref{eq:RL}), of right and left
polarization implemented in this frame leads to ${\cal E}_K=0$
because $k_z^\prime=0$. That is, the ``circularly polarized beam"
becomes a TM mode whose electric field points in the $z$
direction. On the other hand the definition Eq.~(\ref{eq:pm})
leads to an electric field with a projection in the XY plane that
is a superposition of two opposite circular vectors with
amplitudes proportional to $J_{m \mp 1}$, while the $z$ component
of the field is proportional to $J_m$ and has a relative phase
$\pm\pi/2$.

 In Fig.~(1), we illustrate the electric and magnetic
fields in the plane perpendicular to the $z$ axis, as seen in the
antiparaxial frame, for a circularly polarized mode in the sense
of Eq.~(\ref{eq:pm}). Notice that, as previously mentioned,
$\Re{\bf E}$ and $\Re{\bf B}$ are not perpendicular in general.
Edge phase dislocations, saddle points and vortices \cite{nye} are
present. Some of these structures are formed around the zeros of
$J_m$ and $J'_m$. In Fig.~(2), the intensity patterns
$\vert\Re{\bf E}\vert^2$, $\vert\Re{\bf B}\vert^2$ and their sum
are illustrated. Notice the complementary space distribution of
the electric and magnetic fields.

As for the dynamical properties of the field, a noteworthy feature
in the new frame is that the $z$ component of the Poynting vector
is
\begin{equation}
\Big(\Re{\bf E}_K^\prime  \times \Re {\bf B}_K^\prime
\Big)_z=\frac{i}{8}({\cal E}_K{\cal B}_K^* - {\cal B}_K{\cal
E}_K^*)(J_{m-1}^2- J_{m+1}^2) ,\label{eq:Poynt}
\end{equation}
so that it vanishes for pure TE or TM modes, as expected from the
fact that $k_z^\prime =0$. On the other hand, the local flux of
energy for circularly polarized modes in the sense of
Eq.~(\ref{eq:pm}) is different from zero even in this frame.
Nevertheless, the  integral over whole space of $\big(\Re{\bf
E}_K^\prime  \times \Re {\bf B}_K^\prime \big)_z$ is zero since
\begin{equation}
\int_0^\infty [J_{m+1}^2( k_\bot\rho)- J_{m-1}^2(k_\bot\rho)]\rho d\rho =
\frac{4m}{k_\bot}\int_0^\infty J^\prime_m( k_\bot\rho)J_m( k_\bot\rho)d\rho =
0.\label{eq:complete}
\end{equation}

Let us now turn our attention to the angular momentum. Its total
density is defined as
\begin{equation}
{\cal  J} =\frac{1}{4\pi} {\bf r} \times \Big( {\bf E} \times {\bf
B}\Big),
\end{equation}
and it is known that in the Coulomb gauge, up to a surface term,
it can be decomposed into the sum of the so-called {\it orbital}
angular momentum density \cite{jh,mandel}
\begin{equation}
{\cal  L} =\frac{1}{4\pi} \sum_i  {\bf E}_i({\bf r}\times {\bf
\nabla}){\bf A}_i \label{am}
\end{equation}
and the {\it spin} angular momentum density
\begin{equation}
{\cal S} = \frac{1}{4\pi} {\bf E}\times {\bf A} \label{sam}.
\end{equation}

The above equations can be obtained in a relativistically covariant form. Since the
conservation of angular momentum is related to a rotational symmetry, it is natural
to use the Killing vector $k^{\alpha}$ associated to rotations around the $z$ axis.
This vector is defined as $k^{\alpha}\partial_{\alpha} =
\partial/\partial \phi $, and has the standard property
$\nabla_{\alpha} k_{\beta}= - \nabla_{\beta}k_{\alpha}$, where
$\nabla_{\alpha}$ is the covariant derivative (see, e. g.,
Weinberg \cite{weinberg}). In cartesian coordinates ($x^{\mu},
\mu=0$ to 3):
\begin{equation}
k^{\alpha}=(0,-y, x,0)~~,~~~~~\nabla_{\alpha} k_{\beta}=
\epsilon_{0{\alpha}{\beta}3}~.
\end{equation}

Using the energy-momentum tensor for the electromagnetic field
$F_{\mu \nu}$:
\begin{equation}
T^{\alpha \beta} = \frac{1}{4\pi} \Big[ F^{\alpha \mu} F^{\beta
~\cdot}_{~~~\mu} - \frac{1}{4} g^{\alpha \beta} (F^{\lambda \mu}
F_{\lambda \mu}) \Big],
\end{equation}
it follows that the condition $\nabla_{\beta} T^{\alpha \beta} =
0$ implies that the four-vector $J^{\alpha} \equiv k_{\beta}
T^{\alpha \beta}$ is conserved, that is: $\nabla_{\alpha}
J^{\alpha} =0$. Thus, using the fact that $F_{\alpha \beta} =
\nabla_{\alpha} A_{\beta} - \nabla_{\beta} A_{\alpha}$ and
$\nabla_{\alpha}F^{\alpha \beta}=0$, it turns out that
$J^{\alpha}$ can be written as:
\begin{equation}
J^{\alpha} = \frac{1}{4\pi} \Big[F^{\alpha \mu} k^{\nu}
\nabla_{\nu} A_{\mu} + F^{\alpha \mu} A^{\nu} \nabla_{\mu} k_{\nu}
- \nabla_{\mu} (F^{\alpha \mu} k^{\nu} A_{\nu}) - \frac{1}{4}
k^{\alpha} (F^{\lambda \mu} F_{\lambda \mu}) \Big].
\end{equation}
Since $J^{\alpha}$ has zero divergence, the integral over a
three-dimensional hypersurface with normal unit four-vector
$n_{\alpha}$ and volume element $dV$,
\begin{equation}
J \equiv \int dV n_{\alpha} J^{\alpha},
\end{equation}
is independent of the particular choice of the hypersurface and
thus is a conserved quantity. In particular, choosing such
hypersurface as $t=$ constant, which implies that $n_{\alpha}= (1,
{\bf 0})$, it follows that
\begin{equation}
n_{\alpha} J^{\alpha} = \frac{1}{4\pi} \Big[F^{0 m} k^{\nu}
\nabla_{\nu} A_{m} + F^{0 \mu} A^{\nu} \nabla_{\mu} k_{\nu} -
\nabla_{m} (F^{0m} k^{\nu} A_{\nu}) \Big].
\end{equation}
We readily identify the first term in this equation as the orbital
angular momentum density, given by Eq.~(\ref{am}) and the second
term as the spin density, Eq.~(\ref{sam}). As for the third term,
its volume integral involves a three-dimensional divergence and
can be taken as zero if the field vanishes at infinity. Thus, we
have recovered the standard formulas (\ref{am}) and (\ref{sam}) in
a covariant form. The dependence of the reference frame appears
through the choice of time-like unit vector $n_{\alpha}$.

Applying Eq.~(\ref{am}) and taking only the real parts of the
complex fields, it turns out that in the laboratory frame
\begin{eqnarray}
\bar{\cal L}_z ^K &=&\frac{m}{16\pi\omega k_\bot^2} \Big\{\Big(\omega^2 \vert {\cal
E}_K\vert^2 +k_z^2\vert{\cal
B}_K\vert^2\Big)\Big(J_{m-1}^2 +J_{m+1}^2\Big)\nonumber\\
&+& 2\Big[\omega^2 \vert {\cal E}_K\vert^2 \cos 2(\varphi +\varphi_E)
+k_z^2\vert{\cal
B}_K\vert^2\cos 2(\varphi +\varphi_B)\Big]J_{m-1}J_{m+1}\nonumber\\
 &+&2k_z\omega\vert{\cal
E}_K\vert\vert{\cal B}_K\vert \sin(\varphi_{\cal B}-\varphi_{\cal E})\big(J_{m-1}^2 -
J_{m+1}^2\big)\nonumber\\
&+& 2 k_\bot^2 \vert{\cal B}_K\vert^2  [1+\cos 2(\varphi +\varphi_B)]J_{m}^2 \Big\}~,
\end{eqnarray}
where $\varphi =m\phi + k_z z-\omega t$, and $\varphi_{\cal E}$
and $\varphi_{\cal B}$ are the phases of the complex amplitudes
${\cal E}_K$ and ${\cal B}_K$. The density $\bar{\cal L}_z^K$
averaged over a cycle is accordingly
\begin{eqnarray}
\bar{\cal L}_z ^K &=&\frac{m}{16\pi\omega k_\bot^2} \Big[\Big(\omega^2 \vert {\cal
E}_K\vert^2 +k_z^2\vert{\cal
B}_K\vert^2\Big)\Big(J_{m-1}^2 +J_{m+1}^2\Big)\nonumber\\
 &+& 2 k_z\omega \vert{\cal
E}_K\vert\vert{\cal B}_K\vert\sin(\varphi_B-\varphi_{\cal E} )\big(J_{m-1}^2 -
J_{m+1}^2\big)\nonumber\\
&+& 2 k_\bot^2\vert{\cal B}_K\vert^2 J_{m}^2  \Big] ,
\end{eqnarray}
which is non null for all kinds of polarization and is
proportional to the azimuthal number $m$ as expected. As for the
spin density of the electromagnetic field, it follows from the
previous equations that
\begin{eqnarray}
{\cal S}_z^K &=& \frac{1}{16\pi \omega k_\bot^2}\Big[\Big(\omega^2 \vert {\cal
E}_K\vert^2 +k_z^2\vert{\cal
B}_K\vert^2\Big) \big(J_{m+1}^2- J_{m-1}^2\big)\nonumber\\
&+& 2 k_z\omega\vert {\cal E}_K\vert \vert {\cal B}_K\vert\sin(\varphi_B -
\varphi_{\cal E}) \big(J_{m+1}^2+ J_{m-1}^2\big)\Big]~.
\end{eqnarray}
Thus, {\it  in the antiparaxial frame} a purely transverse
magnetic mode has null  ${\cal S}_z^K $, while any other
cylindrical mode, such as a circularly polarized wave, has a
``spin" density in the $z$-direction. Notice, however, that the
integral of the $z$ component of the Poynting vector,
Eq.~(\ref{eq:Poynt}), and the integral of  ${\cal S}_z^K$ over the
whole space are zero in this frame. This result is consistent with
the interpretation of the integral of  ${\cal S}_z^K $ as a
helicity operator. As for the sum of the orbital and spin angular
momentum densities averaged over a cycle, it turns out to be
\begin{equation}
\bar{\cal  J}_z ^K =\frac{1}{16\pi k_\bot} \Big\{ \vert {\cal
E}_K\vert^2 \Big[(m-1)J_{m-1}^2 +(m+1)J_{m+1}^2\Big]+ \vert{\cal
B}_K\vert^2 mJ_{m}^2 \Big\}
\end{equation}
in the antiparaxial frame.

An alternative definition of the orbital and spin angular momentum
densities that has the advantage of being gauge independent has
been proposed by Barnett\cite{barnett}, who defined the
time-averaged z-component of the orbital angular momentum flux
(for complex electric ${\bf E}$ and magnetic ${\bf B}$) as
\begin{equation}
M_{zz}^{ORB}=\frac{1}{16\pi \omega}\Re
\Big[-i\Big(-B_x^*\frac{\partial
E_y}{\partial\phi}+E_y\frac{\partial B^*_x}{\partial\phi}
-E_x\frac{\partial B^*_y}{\partial\phi} +B_y^*\frac{\partial
E_x}{\partial\phi}\Big)\Big],
\end{equation}
and the spin flux as
\begin{equation}
M^{SPIN}_{zz} = \frac{1}{8\pi
\omega}\Re\Big[-i(E_xB_x^*+E_yB_y^*)\Big].
\end{equation}
Applying these formulas to a Bessel mode we find that
\begin{eqnarray}
M_{zz}^{ORB} =&-&\frac{k_z}{16\pi k_\bot^2}\Big(\vert{\cal E}_K\vert^2 +\vert{\cal
B}_K\vert^2\Big)\big[(m-1)J_{m-1}^2+
(m+1)J_{m+1}^2\Big] \nonumber\\
&+&\frac{1}{16\pi\omega}\Big[2(k_z /k_\bot)^2 +1\Big]\vert{\cal E}_K\vert \vert{\cal
B}_K\vert\cdot\nonumber\\&\cdot& \sin(\varphi_E -\varphi_B)\big[(m-1)J_{m-1}^2
-(m+1)J_{m+1}\big] ,
\end{eqnarray}
and
\begin{eqnarray}
M_{zz}^{SPIN}&=&\frac{k_z}{16 \pi k_\bot^2}(\vert {\cal
E}_K\vert^2 +\vert
{\cal B}_K\vert^2)(J_{m-1}^2-J^2_{m+1})\nonumber\\
&+&\frac{1}{16\pi\omega}\Big[2(k_z /k_\bot)^2 +1\Big]\vert{\cal E}_K\vert \vert{\cal
B}_K\vert\nonumber\\
&\cdot&\sin(\varphi_E -\varphi_B)\big(J_{m-1}^2+J^2_{m+1}\big) .
\end{eqnarray}
Thus, according to this definition, the total density of angular
momentum averaged over a cycle is
\begin{eqnarray}
M_{zz}&=&M_{zz}^{ORB} + M_{zz}^{SPIN}\nonumber\\
&=&-\frac{mk_z}{16\pi k_\bot^2}\Big(\vert{\cal E}_K\vert^2
+\vert{\cal B}_K\vert^2\Big)\big(J_{m-1}^2+J_{m+1}^2\Big) \nonumber\\
&&+\frac{m}{16\pi\omega}\Big[2(k_z /k_\bot)^2 +1\Big]\vert{\cal
E}_K\vert \vert{\cal B}_K\vert \sin(\varphi_E
-\varphi_B)\big(J_{m-1}^2 -J_{m+1}\big) ~.
\end{eqnarray}
As a consequence, {\it in the antiparaxial frame}, both
$M_{zz}^{ORB}$ and $M_{zz}^{SPIN}$ vanish for pure TE or TM modes,
but not for circular modes. This apparent inconsistency may be due
to the fact that the total $M_{zz}$ transforms as $M_{zz}
\rightarrow M_{zz} + \gamma v J_z$ under a Lorentz boost along the
$z$ axis, unlike $J_z$ which is invariant, so that its value
depends on a given reference frame moving along the beam.

In conclusion, a relativistic analysis reveals the existence of a particular moving
frame in which Bessel beams have a simpler form. In practice, this may be a frame
moving at relativistic speed. However, since we are dealing with exact solutions of
Maxwell's equations, some invariants can be easily calculated and the covariance of
variables such as the angular momentum densities can be studied. Although the
definitions of the so called orbital and angular momentum densities have some
ambiguities, it must be expected that these quantities do not change qualitatively,
or even vanish, by a Lorentz transformation. Actually, we have shown that there are
important cancellations in the antiparaxial limit. Thus it seems more appropriate to
interpret the standard angular momentum density ${\cal S}_z^K$ as the helicity
density, in accordance with our discussions in a previous paper\cite{jh}. Moreover,
the orbital angular momentum is expected to be finite in the antiparaxial frame if it
is different to zero in any other frame, but that is not the case when the expression
of $M_{zz}^{ORB}$ is applied to TE or TM modes. It is also clear from our analysis
that the definition of polarization which is used for plane waves cannot be applied
unambiguously to a Bessel beam. One must be careful in defining what is meant by
polarized states: in fact, we have shown two different definitions leading to
different properties. As a further application of the present relativistic analysis,
we will study the motion of charged particles in the field of Bessel beams in a
forthcoming paper.

\section{Figures captions}
Figure 1. Magnetic and electric fields in the plane perpendicular
to the $z$ axis, as seen in the antiparaxial frame, for a
circularly polarized mode [defined by Eq.~(\ref{eq:pm})] with
$m=2$.

\noindent Figure 2. Intensity patterns $\vert\Re{\bf E}\vert^2$,
$\vert\Re{\bf B}\vert^2$ and their sum, as seen in the
antiparaxial frame, for a circularly polarized mode [defined by
Eq.~(\ref{eq:pm})] with $m=2$.

\end{document}